\documentclass{article}
\usepackage{spconf,amsmath,graphicx}
\usepackage{comment,psfrag,amssymb,booktabs,tabularx,pgfplots} 

\title{Motion Estimation for Fisheye Video Sequences Combining Perspective Projection with Camera Calibration Information}
\name{Andrea Eichenseer, Michel B\"atz, and Andr\'e Kaup\thanks{This work was supported by the Research Training Group 1773 “Heterogeneous Image Systems”, funded by the German Research Foundation (DFG).}}
\address{Multimedia Communications and Signal Processing\\
	Friedrich-Alexander University Erlangen-N\"urnberg (FAU), Cauerstr. 7, 91058 Erlangen, Germany\\
	}
\begin{document}
\ninept 
\maketitle
\begin{abstract}
Fisheye cameras prove a convenient means in surveillance and automotive applications as they provide a very wide field of view for capturing their surroundings.
Contrary to typical rectilinear imagery, however, fisheye video sequences follow a different mapping from the world coordinates to the image plane which is not considered in standard video processing techniques.
In this paper, we present a motion estimation method for real-world fisheye videos by combining perspective projection with knowledge about the underlying fisheye projection.
The latter is obtained by camera calibration since actual lenses rarely follow exact models.
Furthermore, we introduce a re-mapping for ultra-wide angles which would otherwise lead to wrong motion compensation results for the fisheye boundary.
Both concepts extend an existing hybrid motion estimation method for equisolid fisheye video sequences that decides between traditional and fisheye block matching in a block-based manner.
Compared to that method, the proposed calibration and re-mapping extensions yield gains of up to 0.58 dB in luminance PSNR for real-world fisheye video sequences.
Overall gains amount to up to 3.32 dB compared to traditional block matching.

\end{abstract}
\begin{keywords}
Fisheye Lens, Camera Calibration, Perspective Projection, Motion Estimation, Motion Compensation
\end{keywords}
\section{Introduction}
\label{sec:intro}

\begin{figure}[t]
\centering
\centerline{\includegraphics[width=\columnwidth]{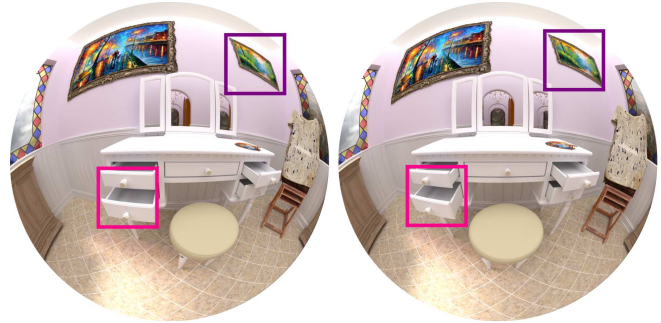}}
\vspace{-0.3cm}
\caption{Block matching between two fisheye images. For visualization purposes, the motion between current (right) and reference (left) frame has been exaggerated.}
\label{fig:blockmatchingexmp}
\vspace{-0.5cm}
\end{figure}

Fisheye cameras prove useful in any kind of situation where a large field of view (FOV) is desired or necessary.
Applications can range from automotive~\cite{gehrig, auto1, auto2, compvis} over video surveillance~\cite{surveillance, surveillance2} to image-based virtual reality~\cite{vr1,vr2}, thus creating quite a number of potential image and video processing tasks.
Be it compression and coding for transmission or post-processing in the form of image enhancement or error concealment~\cite{eichenseer2015tecfish}, all those tasks face a certain challenge when having to deal with fisheye video sequences.
This challenge arises due to the non-rectilinear or non-perspective nature of fisheye images.
Since fisheye lenses~\cite{miyamoto1964fel} are designed in a way that allows capturing FOVs of more than 180$^\circ$, the resulting fisheye images are radially distorted, meaning that lines which should be straight are actually depicted as arcs.
As a consequence, even a simple translational motion of the scene or camera does not correspond to just a translation of the single fisheye images, which in turn creates problems for algorithms that are based on a translational motion model.
A visualization of this is given in Fig.~\ref{fig:blockmatchingexmp} which shows two fisheye frames of a video sequence where the camera movement was purely translational in horizontal direction.
The highlighted squares show that the image content is not only translated but also appears warped so that traditional block matching will not find a good predictor.
To mitigate this problem, motion estimation techniques especially adapted to fisheye imagery have been proposed in the literature.

In~\cite{jin2015warpedfisheye}, the authors argue that neighboring motion vectors will differ in fisheye videos even when dealing with global motion.
Thus, they determine the difference between neighboring motion vectors, which would be equal in perspective images, and thus compensate for the fisheye characteristics.
They demonstrate their method for perspective videos which were warped with an equidistant fisheye model.
In~\cite{eichenseer2015motionfish}, a hybrid technique combines traditional block matching with a fisheye variant that projects the fisheye pixel positions to their perspective representation by means of the equisolid fisheye model as well as the pinhole model. The latter can be used as an approximation for basically any kind of perspective imagery.
Motion estimation is thus conducted on perspective coordinates and the translational motion model is valid.
Transferring the shifted coordinates back into the fisheye representation then allows for a fisheye motion compensation that does not require distortion corrected images and preserves the original round form of the video frames.
The method was shown to work for synthetic equisolid fisheye sequences rendered from 3D scenes as well as for real-world fisheye videos captured with an actual fisheye camera.

In this paper, we build upon the hybrid motion estimation technique introduced in~\cite{eichenseer2015motionfish}.
Contrary to the original contribution, 
we now propose to use calibration information instead of the equisolid model.
This especially makes sense when dealing with real-world sequences, as actual fisheye lenses rarely follow an exact trigonometrical function.
Furthermore, we deal with ultra-wide incident angles beyond 90$^\circ$ as they cause the pinhole model to fail and thus we create a better prediction at the boundary of the fisheye image.

In the following, we provide a brief overview of traditional block-based motion estimation as well as hybrid motion estimation for fisheye video sequences in Section~\ref{sec:stateoftheart}. In Section~\ref{sec:calib}, the proposed calibrated motion estimation is described along with the re-mapping for ultra-wide angles. Section~\ref{sec:results} provides the simulation results for both extensions and Section~\ref{sec:conclusion} concludes this paper.

\section{Motion Estimation for Fisheye Video}
\label{sec:stateoftheart}

The single frames of a typical video sequence are highly correlated in the temporal direction, which means that the less pronounced the global or local motion of a scene is, the more similar two subsequent video frames are.
This characteristic is, for example, exploited in video coding in the form of temporal prediction, but estimating the motion between two frames is a step that must be completed in many signal processing tasks as well.
A simple yet reliable way to predict the current frame from its temporal neighbor can be realized by a block matching algorithm~\cite{blockmatching}.
Such a block-based algorithm matches the current block to be predicted to each block within a defined search range in the reference frame.
Minimizing the residual error using a metric like the sum of squared differences (SSD) then yields the optimum reference block to be used for prediction or motion compensation.

This method works well in conventional perspective images with translation as the predominant kind of motion.
In fisheye videos, traditional block matching is not able to find shifted versions of the current block since the translation caused the scene not only to shift but also to bend or seemingly warp (cf. Fig.~\ref{fig:blockmatchingexmp}).
To compensate for this, a block can be interpreted as a set of pixel coordinates which are projected to a perspective representation using the underlying fisheye model.
The perspective coordinates are then shifted by a motion vector candidate and subsequently warped back to their new fisheye representations.
For the compensation, the new pixel coordinates are then used to extract the corresponding luminance values from the reference image.
Note that it is not necessary to perform an actual distortion correction that would yield a perspective image.
This summarizes the method introduced in~\cite{eichenseer2015motionfish}.
In the following, we will adapt this method by using camera calibration instead of an equisolid fisheye model since in a realistic scenario, the fisheye projection function is unknown and not trigonometrical.
We will furthermore include a re-mapping for ultra-wide angles (FOV $>$ 180$^\circ$) to compensate for an otherwise failing pinhole model.

\section{Fisheye Motion Estimation Using Calibration Information}
\label{sec:calib}

\subsubsection*{Perspective Projection and Calibrated Re-Projection}
\vspace{-0.2cm}

Since actual fisheye lenses rarely follow an exact projection function such as the equisolid one~\cite{miyamoto1964fel}, a calibration is necessary to obtain information about the camera parameters and the mapping employed.
Replacing the exact equisolid mapping used in~\cite{eichenseer2015motionfish} by the projection function obtained via calibration, a more accurate motion estimation can be performed for real-world sequences taken with an actual fisheye camera.
Depending on the method employed, a sequence of images showing a calibration pattern might be necessary to conduct the calibration.


We assume that the used camera calibration method yields the information on the projection function in the form of a polynomial $p(\theta)$, where $\theta$ denotes the angle of the incident ray of light measured against the optical axis.
This polynomial
\vspace{-0.25cm}
\begin{equation}
\vspace{-0.25cm}
p(\theta) = a_n\theta^n + a_{n-1}\theta^{n-1} + \ldots  + a_1\theta + a_0 = \sum\limits_{i=0}^n a_i\theta^i
\end{equation}
can be of any specified order $n$, which should be chosen according to the calibration specifications.
For the calibration toolbox we used, for example, a 4th-order polynomial is recommended as it is said to yield the best results~\cite{scara2}.
Having obtained knowledge about the projection used by the fisheye camera, this knowledge can be included in the fisheye motion estimation.

In order to project the pixel coordinates of the current fisheye image block to perspective coordinates, the pinhole model is used as this is the model that describes perspective projection:
\vspace{-0.1cm}
\begin{equation}
\label{eq:ftop}
\vspace{-0.1cm}
r_{\mathrm{p}} = f \tan \left(\theta\right), \quad\mathrm{with}\: \theta = p^{-1}\left(r_{\mathrm{f}}\right).
\end{equation}
The focal length $f$ is given by the lens or estimated via calibration.
$r_{\mathrm{p}}$ describes the distance to the perspective image center and is synonymously referred to as (perspective) radius.
The angle $\theta$ is given by the inverse polynomial, where $r_{\mathrm{f}}$ describes the radius of the fisheye pixel coordinates and can be obtained by a Cartesian-to-polar transform.
Note that the polar coordinates are centered around the calibrated image center as opposed to the center defined by the image dimensions.

Shifting the pixel coordinates by the motion vector candidate $\boldsymbol{m}$ is then conducted with the perspective coordinates.
Afterward, the new fisheye coordinates can be re-obtained from the shifted perspective coordinates by using the regular polynomial as this actually describes the mapping to the fisheye image:
\vspace{-0.2cm}
\begin{equation}
\label{eq:ptof}
\vspace{-0.2cm}
r_{\mathrm{f},\boldsymbol{m}} = p\left(\theta\right), \quad\mathrm{with}\: \theta = \arctan\left(\frac{r_{\mathrm{p},\boldsymbol{m}}}{f}\right).
\end{equation}
This time, the angle $\theta$ is obtained from the inverted pinhole model.
The fisheye coordinates are then used to extract the corresponding luminance values from the upsampled and interpolated reference frame, thus creating a motion compensated block.

This procedure is repeated until the best motion vector candidate and thus the best compensated block is found for the current block.
As a criterion for minimizing the residual error, we use the sum of squared differences (SSD).
More details on the inclusion of the projections, the hybrid implementation as well as the compensation procedure can be found in~\cite{eichenseer2015motionfish} and~\cite{eichenseer2015tecfish}.

\vspace{-0.1cm}
\subsubsection*{Compensating for Ultra-Wide Angles}
\vspace{-0.2cm}

\begin{figure}[t]
\centering
\centerline{\includegraphics[width=\columnwidth]{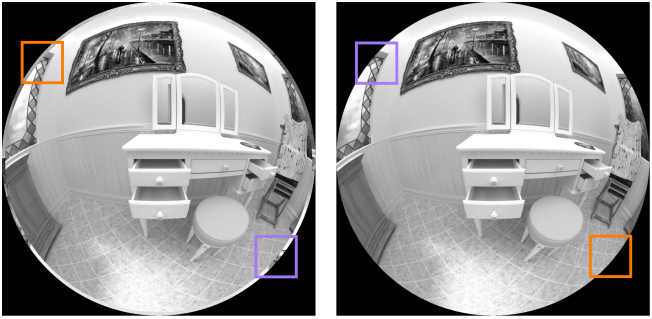}}
\vspace{-0.4cm}
\caption{Comparison of non-hybrid fisheye motion compensation with (right, EME+) and without (left, EME) additionally compensating for $\theta>\pi/2$. The underlying FOV corresponds to 185$^\circ$.}
\label{fig:borderexmp}
\vspace{-0.5cm}
\end{figure}

Important to note is the tangent function involved in the fisheye-to-perspective projection~(\ref{eq:ftop}).
For all $\theta > \pi/2$, which occur for an FOV of more than 180$^\circ$, a negative value is returned for the radius.
Leaving this untreated makes the fisheye motion compensation fail for these angles as the wrong image content is mapped into the corresponding region.
Using the hybrid method combining fisheye motion estimation with traditional block matching~\cite{eichenseer2015motionfish}, the traditional path would be chosen in this case.
While we still retain a hybrid motion estimation technique in our simulations, we also compensate for the wrong mapping caused by ultra-wide angles. The first step consists of inverting the sign of the motion vector candidate $\boldsymbol{m}$ 
for all coordinates that were assigned a negative value for the radius:
\vspace{-0.2cm}
\begin{equation}
\vspace{-0.2cm}
\boldsymbol{m} = 
\begin{cases}
(-\Delta x_{\mathrm{p}}, -\Delta y_{\mathrm{p}}), & \forall\: \{(x_{\mathrm{p}},y_{\mathrm{p}}) \:|\: r_{\mathrm{p}} < 0\} \\
(\Delta x_{\mathrm{p}}, \Delta y_{\mathrm{p}}), & \text{otherwise}.
\end{cases}
\end{equation}
Since the motion vector candidate $(\Delta x_{\mathrm{p}}, \Delta y_{\mathrm{p}})$ is defined in the Cartesian coordinate system, $(x_{\mathrm{p}},y_{\mathrm{p}})$ denotes the Cartesian representation belonging to the polar coordinates $(r_{\mathrm{p}},\phi_{\mathrm{p}})$.
As the second step, the angle $\phi_{\mathrm{p},\boldsymbol{m}}$, obtained after adding $\boldsymbol{m}$, has to be adjusted:
\vspace{-0.1cm}
\begin{equation}
\vspace{-0.1cm}
\phi'_{\mathrm{p},\boldsymbol{m}} = 
\phi_{\mathrm{p},\boldsymbol{m}} - \pi, \quad \forall\: r_{\mathrm{p}} < 0\:.
\end{equation}
After projecting back the shifted and manipulated perspective coordinates $(r_{\mathrm{p},\boldsymbol{m}},\phi'_{\mathrm{p},\boldsymbol{m}})$ to their fisheye representation $(r_{\mathrm{f},\boldsymbol{m}},\phi'_{\mathrm{f},\boldsymbol{m}})$, 
the third step consists of:
\vspace{-0.1cm}
\begin{equation}
\vspace{-0.05cm}
r'_{\mathrm{f},\boldsymbol{m}} = 
r_{\mathrm{f},\boldsymbol{m}} + 2(r_{\mathrm{f},\pi} - r_{\mathrm{f},\boldsymbol{m}}), \quad \forall\:  r_{\mathrm{p}} < 0\:,
\end{equation}
with $r_{\mathrm{f},\pi} = p(\pi/2)$ describing the radius obtained for $\theta=\pi/2$, i.\,e., an FOV of 180$^\circ$ which is equal to $\pi$.
That way, a fisheye image is obtained that again contains proper content for $\theta > \pi/2$.
Fig.~\ref{fig:borderexmp} visualizes this problem for an example frame and is best viewed enlarged on a screen.
The left image uses equisolid fisheye motion estimation (EME) which is part of the original hybrid motion estimation method, while the right image uses the extended version (EME+) which includes the coordinate re-mapping for $\theta>\pi/2$.

Note that the proposed calibrated method as well as the compensation for ultra-wide angles just introduced is purely based on coordinates and does not rely on image distortion correction or using actual perspective images. Thus, the fisheye sequences are preserved in their original form.

\section{Simulation Setup and Results}
\label{sec:results}

In this section, the following nomenclature is used:
TME describes traditional block-based motion estimation; HME is the original hybrid method of~\cite{eichenseer2015motionfish} and combines TME with the equisolid fisheye motion estimation EME.
EME+ is the extended version of EME including the above-mentioned treatment for $\theta > \pi/2$ and thus forms HME+ when combined with TME.
CME and CHME are the calibrated versions of EME and HME, whereas CME+ and CHME+ again include the compensation for very wide angles $\theta$.

\begin{figure}[t]
\centering
\input{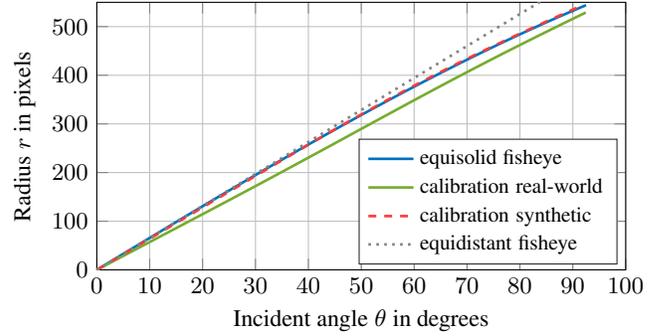}
\vspace{-0.8cm}
\caption{Calibration results compared to the equisolid fisheye model.}
\vspace{-0.5cm}
\label{fig:calibration}
\end{figure}

For the proposed calibrated method, we employed a checkerboard pattern, one of the most commonly used calibration patterns, and captured images showing that pattern at different distances, positions, and angles with our fisheye camera.
Since the acquired image set is the basis for obtaining good calibration results, it was ensured that all corners of the checkerboard are clearly visible in all used images.
Please note that we used the exact same camera (FOV$=$185$^\circ$) and settings for both capturing the fisheye sequences and the calibration images.
While automatic corner extraction tools exist~\cite{scara3}, we chose to extract the checkerboard corners manually for the sake of accuracy.
For the actual calibration, we used the OCamCalib toolbox~\cite{scara2} and thus obtained the projection function corresponding to our fisheye camera.
Fig.~\ref{fig:calibration} shows the calibration result as a function of the radius $r$, which corresponds to the distance to the image center, over the angle $\theta$ of the incident angle of light. The green curve thus corresponds to the polynomial $r_{\mathrm{f}}  = p(\theta)$.
The equisolid angle projection is shown for comparison purposes in blue.
Evidently, the fisheye camera's underlying model is not an exactly equisolid one.
While the equisolid projection function describes a sine ($r = 2f\sin(\theta/2)$), the calibration result looks to be of a more linear nature without being exactly linear.
Please note that the calibration result does not match the linear equidistant fisheye model which is not regarded in our method.

\begin{table*}[t]
\small
\caption{Average luminance PSNR results and gain of both original HME and proposed CHME+ over traditional block matching in dB for two block sizes: 16$\times$16 pixels (top half) and 32$\times$32 pixels (bottom half). A search range of 64 pixels was employed for all tests.}
\label{tab:psnrcomp}
\centering
\renewcommand\arraystretch{0.9}
\begin{tabularx}{\textwidth}{ll|c|cc|cc|cc|cc|cc}
\toprule
Sequence & Frames & TME & EME & HME & EME+ & HME+& CME & CHME & CME+ & CHME+ & \hspace{-0.1cm}$\Delta$HME & \hspace{-0.3cm}$\Delta$CHME+\\
\midrule
\textit{CheckercubeB} & 31--50 & 42.28 & 18.42 & 44.23 & 31.49 & \textbf{44.27} & 19.43 & 44.23 & 31.55 & 44.26 & 1.95 & 1.98\\
\textit{HallwayD} & 31--50 & 34.75 & 24.54 & 38.39 & 37.98 & \textbf{38.55} & 25.39 & 38.28 & 38.17 & 38.49 & 3.64 & 3.74\\
\addlinespace
\textit{AlfaA} & 251--280 & 34.99 & 32.58 & 37.96 & 37.81 & 37.97 & 28.78 & 38.17 & 38.09 & \textbf{38.20} & 2.97 & 3.21\\
\textit{TestchartB} & 201--230 & 36.76 & 33.09 & 38.84 & 38.66 & 38.86 & 31.15 & 38.99 & 38.89 & \textbf{39.03} & 2.08 & 2.27\\
\textit{ClutterB} & 251--280 & 36.53 & 25.27 & 37.64 & 37.23 & 37.66 & 22.99 & 37.83 & 37.39 & \textbf{37.89} & 1.10 & 1.36\\
\textit{LectureB} & 288--317 & 34.78 & 20.19 & 36.22 & 35.58 & 36.32 & 16.80 & 36.20 & 35.79 & \textbf{36.50} & 1.44 & 1.72\\
\textit{LibraryA} & 191--210& 35.94 & 22.31 & 37.27 & 31.83 & 37.27 & 20.95 & 37.31 & 31.01 & \textbf{37.33} & 1.33 & 1.39\\
\textit{LibraryB} & 126--155 & 37.43 & 20.85 & 39.35 & 38.94 & 39.36 & 18.24 & 39.73 & 39.59 & \textbf{39.79} & 1.92 & 2.36\\
\textit{LibraryD} & 101--130 & 36.54 & 20.65 & 38.11 & 38.08 & 38.13 & 17.63 & 38.59 & 38.64 & \textbf{38.67} & 1.57 & 2.13\\
\midrule
\textit{CheckercubeB} & 31--50 & 38.82 & 18.32 & 40.35 & 30.16 & \textbf{40.38} & 19.32 & 40.30 & 30.20 & 40.33 & 1.53 & 1.51\\
\textit{HallwayD} & 31--50 & 34.12 & 24.42 & 37.40 & 37.01 & \textbf{37.63} & 25.28 & 37.27 & 37.22 & 37.57 & 3.28 & 3.45\\
\addlinespace
\textit{AlfaA} & 251--280 & 34.59 & 32.31 & 37.66 & 37.51 & 37.67 & 28.58 & 37.89 & 37.81 & \textbf{37.91} & 3.07 & 3.32\\
\textit{TestchartB} & 201--230 & 35.75 & 32.69 & 38.10 & 37.90 & 38.11 & 30.88 & 38.30 & 38.19 & \textbf{38.33} & 2.35 & 2.57\\
\textit{ClutterB} & 251--280 & 36.05 & 25.13 & 37.06 & 36.72 & 37.11 & 22.66 & 37.23 & 36.87 & \textbf{37.31} & 1.00 & 1.26\\
\textit{LectureB} & 288--317 & 33.96 & 20.02 & 35.38 & 34.91 & 35.54 & 16.75 & 35.38 & 35.16 & \textbf{35.72} & 1.42 & 1.77\\
\textit{LibraryA} & 191--210& 35.18 & 22.10 & 36.23 & 31.26 & 36.23 & 20.67 & 36.28 & 30.52 & \textbf{36.30} & 1.04 & 1.11\\
\textit{LibraryB} & 126--155 & 37.08 & 20.79 & 39.07 & 38.67 & 39.08 & 18.17 & 38.46 & 39.32 & \textbf{39.51} & 1.99 & 2.43\\
\textit{LibraryD} & 101--130 & 36.23 & 20.58 & 37.84 & 37.83 & 37.87 & 17.55 & 38.35 & 38.41 & \textbf{38.42} & 1.61 & 2.19\\
\bottomrule
\end{tabularx}
\vspace{-0.33cm}
\end{table*}

\begin{figure*}[t]
\centering
\psfrag{a1}[cB][ct]{\small HME, 37.41 dB}
\psfrag{a2}[cB][ct]{\small HME+, 37.42 dB}
\psfrag{a3}[cB][ct]{\small CHME, 38.29 dB}
\psfrag{a4}[cB][ct]{\small CHME+, 38.34 dB}
\centerline{\includegraphics[width=\textwidth]{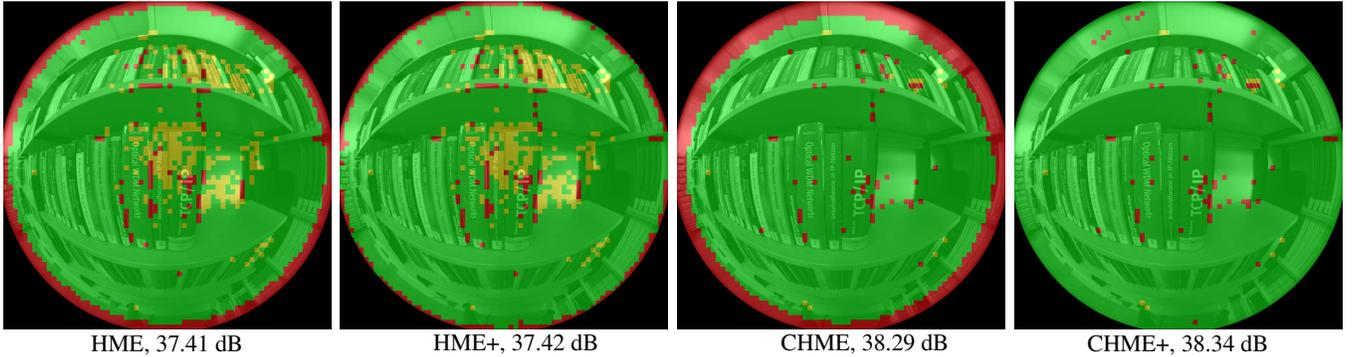}}
\vspace{-0.8cm}
\caption{Hybrid compensation overlays for frame 101 of sequence \textit{LibraryD} as returned by the original HME, HME+ including wide-angle compensation, CHME using calibration, and CHME+ combining both extensions. The corresponding result for TME amounts to 35.85 dB.}
\label{fig:overlays}
\vspace{-0.5cm}
\end{figure*}

For verifying the proposed method, we first made use of some synthetic sequences (\textit{CheckercubeB} and \textit{HallwayD}) that exhibit purely translational motion.
Those sequences are part of the fisheye data set introduced in~\cite{eichenseer2016dataset} and rendered according to the equisolid fisheye model, thus following the same projection function as the hybrid fisheye motion estimation HME proposed in~\cite{eichenseer2015motionfish}.
To include calibration information, we used the \textit{CheckercubeC} sequence of the same data set since it shows a checkerboard pattern.
We again used the OCamCalib toolbox to obtain the needed projection function.
The result of this synthetic calibration is shown in Fig.~\ref{fig:calibration} as the dashed red line.
It fits the equisolid function quite well so that it can be concluded that an accurate calibration was performed.
In the implementation, the polynomial $p(\theta)$ and its inverse $p^{-1}(r_{\mathrm{f}})$ are realized as a lookup table that relates $\theta$ to $r_{\mathrm{f}}$.

For the actual test on real-world fisheye video material, seven sequences were chosen: \textit{AlfaA} and \textit{TestchartB} show a translational motion of a large object; \textit{ClutterB}, \textit{LectureB}, and \textit{LibraryB} exhibit a global translational camera motion; \textit{LibraryD} contains shaky camera motion, while \textit{LibraryA} shows a forward camera motion (zoom).
These sequences are part of the same data set as the synthetic ones and were cropped from their original 1150$\times$1086 pixels to 1088$\times$1086 pixels to include only the relevant area of the fisheye image.
Details on the sequences, the camera, and the settings used during acquisition can be taken from~\cite{eichenseer2016dataset} or the corresponding website.

Table~\ref{tab:psnrcomp} summarizes the average luminance PSNR results for the regarded sequences and motion compensation techniques when using block sizes of 16$\times$16 and 32$\times$32 pixels.
For each sequence, 20 to 30 consecutive frames have been used for the evaluation.
The PSNR calculation was conducted within the same round compensated fisheye images, excluding the black border, for all methods. 
Since both HME and CHME as well as EME and CME perform about equally well for the synthetic sequences, the verification of the calibration is considered successful.
For the real-world sequences, gains are achieved throughout all tests conducted.
While the compensation for ultra-wide angles does not contribute much to the hybrid methods, the fisheye methods (EME and CME) themselves profit tremendously and gain up to 18~dB (EME+) and 21~dB (CME+) in extreme cases (\textit{LibraryB} and \textit{LibraryD}).
Note that for EME+, we used $r_{\mathrm{f},\pi} = 2f\sin{\left(\pi/4\right)}$ since this method is based on the equisolid fisheye model and not on calibration.
Integrating calibration information into the projection to perspective coordinates and back yield gains of about 0.3~dB on average which can rise to more than 0.5~dB for selected sequences.
As the sequence \textit{LibraryA} does not comply with the translational motion model, the gains achieved here are expectedly low.
The direct comparison of gains obtained by the state-of-the-art HME and the proposed CHME+ compared to traditional block matching TME is also provided in the table.
CHME+ achieves overall gains of well beyond 1~dB with possible maximum gains of more than 3~dB.

Fig.~\ref{fig:overlays} shows the hybrid motion compensation results for all discussed methods, HME, HME+, CHME, and CHME+, overlaid with a colored decision mask that indicates which method, traditional or fisheye, was used for which block.
Green color denotes fisheye motion estimation, while red color denotes traditional block-matching.
Yellow means that both compensated blocks are equal according to the employed error metric SSD.
The effect of re-mapping coordinates for ultra-wide angles is nicely shown by the masks.
Since CHME uses calibration information, the mask is shifted by the calibrated image center.
Even more noticeable is the broader red border. Since for the polynomial, the radius corresponding to $\theta = \pi/2$ (516~pixels) is smaller than the radius obtained by the equisolid model (532~pixels), the problematic area is larger and thus, more blocks get compensated by TME.
This fact is also reflected by the PSNR results which decrease accordingly.
For the synthetic sequences, on the other hand, the corresponding calibrated radius is a little greater (535~pixels), which is why the results, which should be equal to those of the equisolid method, differ slightly.
For some of the real-world sequences, it is almost possible to rely solely on CME+ for motion estimation since the re-mapping for ultra-wide angles leads to only small trade-offs between CHME+ and CME+.
This observation is not true in general, however, which is why further work has to be invested to achieve a non-hybrid solution.



\section{Conclusion}
\label{sec:conclusion}
\vspace{-0.05cm}
In this paper, we adapted a hybrid motion estimation technique for equisolid fisheye video sequences by including camera calibration information to better predict real-world sequences.
We further provided a compensation for ultra-wide incident angles which would otherwise lead to a wrong motion compensation at the boundary of the fisheye images.
Combining both concepts achieved gains of about 0.3 dB compared against the state-of-the-art method, while a big step was taken towards a non-hybrid motion estimation technique. 
Current research includes an in-depth analysis of the two proposed extensions for different types of motion as well as a thorough investigation of the visual results.
Building upon that, an integration of an affine motion model into the estimation process~\cite{narroschke2013affine} to better cope with panning, zoom, rotation, and mixed motion in general, is part of future research.
As the search range is still a limiting factor for fisheye motion estimation, an adaptive motion search that considers the radius will also be investigated.
Ultimately, the hybrid concept is to be replaced by a single method.
Estimating parameters directly from the given images without any prior calibration procedure should further generalize the proposed calibrated fisheye motion estimation method.

\vfill
\pagebreak

\bibliographystyle{IEEEbib}
\bibliography{refs}

\begin{thebibliography}{10}

\bibitem{gehrig}
S.~Gehrig, C.~Rabe, and L.~Kr\"uger,
\newblock ``{6D Vision Goes Fisheye for Intersection Assistance},''
\newblock in {\em Proceedings of the Canadian Conference on Computer and Robot
  Vision}, Windsor, Canada, May 2008, pp. 34--41.

\bibitem{auto1}
C.-H. Kum, D.-C. Cho, M.-S. Ra, and W.-Y. Kim,
\newblock ``{Lane Detection System with Around View Monitoring for Intelligent
  Vehicle},''
\newblock in {\em Proceedings of the Second International SoC Design
  Conference}, Busan, South Korea, November 2013, pp. 215--218.

\bibitem{auto2}
D.~Dooley, B.~McGinley, C.~Hughes, L.~Kilmartin, E.~Jones, and M.~Glavin,
\newblock ``{A Blind-Zone Detection Method Using a Rear-Mounted Fisheye Camera
  With Combination of Vehicle Detection Methods},''
\newblock {\em IEEE Transactions on Intelligent Transportation Systems}, vol.
  17, no. 1, pp. 264--278, January 2016.

\bibitem{compvis}
G.~H. Lee, F.~Faundorfer, and M.~Pollefeys,
\newblock ``{Motion Estimation for Self-Driving Cars with a Generalized
  Camera},''
\newblock in {\em Proceedings of the IEEE Conference on Computer Vision and
  Pattern Recognition}, Portland, Oregon, USA, June 2013, pp. 2746--2753.

\bibitem{surveillance}
M.~Findeisen, L.~Meinel, M.~Hes, A.~Apitzsch, and G.~Hirtz,
\newblock ``{A Fast Approach for Omnidirectional Surveillance with Multiple
  Virtual Perspective Views},''
\newblock in {\em Proceedings of the IEEE International Conference on Computer
  as a Tool (EUROCON)}, Zagreb, Croatia, July 2013, pp. 1578--1585.

\bibitem{surveillance2}
Y.~Kubo, T.~Kitaguchi, and J.~Yamaguchi,
\newblock ``{Human Tracking Using Fisheye Images},''
\newblock in {\em Proceedings of the SICE Annual Conference}, Takamatsu, Japan,
  September 2007, pp. 2013--2017.

\bibitem{vr1}
Y.~Xiong and K.~Turkowski,
\newblock ``{Creating Image-Based VR Using a Self-Calibrating Fisheye Lens},''
\newblock in {\em Proceedings of the IEEE Computer Society Conference on
  Computer Vision and Pattern Recognition}, San Juan, Puerto Rico, June 1997,
  pp. 237--243.

\bibitem{vr2}
C.~Chen, H.~Yang, J.~Fan, Y.~Ding, and C.~Han,
\newblock ``{Robust Contour Extraction of Fisheye Images for Image-Based
  Virtual Reality},''
\newblock in {\em Proceedings of the 3rd International Congress on Image and
  Signal Processing}, Yantai, China, October 2010, vol.~3, pp. 1014--1017.

\bibitem{eichenseer2015tecfish}
A.~Eichenseer, J.~Seiler, M.~B\"atz, and A.~Kaup,
\newblock ``{Temporal Error Concealment for Fisheye Video Sequences Based on
  Equisolid Re-Projection},''
\newblock in {\em Proceedings of the European Signal Processing Conference},
  Nice, France, September 2015, pp. 1636--1640.

\bibitem{miyamoto1964fel}
K.~Miyamoto,
\newblock ``{Fish Eye Lens},''
\newblock {\em Journal of the Optical Society of America}, vol. 54, no. 8, pp.
  1060--1061, August 1964.

\bibitem{jin2015warpedfisheye}
G.~Jin, A.~Saxena, and M.~Budagavi,
\newblock ``{Motion Estimation and Compensation for Fisheye Warped Video},''
\newblock in {\em Proceedings of the IEEE International Conference on Image
  Processing}, Quebec City, Canada, September 2015, pp. 2751--2755.

\bibitem{eichenseer2015motionfish}
A.~Eichenseer, M.~B\"atz, J.~Seiler, and A.~Kaup,
\newblock ``{A Hybrid Motion Estimation Technique for Fisheye Video Sequences
  Based on Equisolid Re-Projection},''
\newblock in {\em Proceedings of the IEEE International Conference on Image
  Processing}, Quebec City, Canada, September 2015, pp. 3565--3569.

\bibitem{blockmatching}
M.~Santamaria and M.~Trujillo,
\newblock ``{A Comparison of Block-Matching Motion Estimation Algorithms},''
\newblock in {\em Proceedings of the IEEE 7th Colombian Computing Congress},
  Medellin, Colombia, October 2012, pp. 1--6.

\bibitem{scara2}
D.~Scaramuzza, A.~Martinelli, and R.~Siegwart,
\newblock ``{A Toolbox for Easily Calibrating Omnidirectional Cameras},''
\newblock in {\em Proceedings of the IEEE/RSJ International Conference on
  Intelligent Robots and Systems}, Beijing, China, October 2006, pp.
  5695--5701.

\bibitem{scara3}
M.~Rufli, D.~Scaramuzza, and R.~Siegwart,
\newblock ``{Automatic Detection of Checkerboards on Blurred and Distorted
  Images},''
\newblock in {\em Proceedings of the IEEE/RSJ International Conference on
  Intelligent Robots and Systems}, Nice, France, September 2008, pp.
  3121--3126.

\bibitem{eichenseer2016dataset}
A.~Eichenseer and A.~Kaup,
\newblock ``{A Data Set Providing Synthetic and Real-World Fisheye Video
  Sequences},''
\newblock in {\em Proceedings of the IEEE International Conference on
  Acoustics, Speech and Signal Processing}, Shanghai, China, March 2016,
\newblock Available online: \texttt{www.lms.lnt.de/fisheyedataset}.

\bibitem{narroschke2013affine}
M.~Narroschke and R.~Swoboda,
\newblock ``{Extending HEVC by an Affine Motion Model},''
\newblock in {\em Proceedings of the Picture Coding Symposium}, San Jose,
  California, USA, December 2013, pp. 321--324.

\end{thebibliography}

\end{document}